\documentclass[aps,prb,twocolumn]{revtex4-1}
\usepackage[latin1]{inputenc}
\usepackage[english]{babel}
\usepackage{amsmath}
\usepackage{amsfonts}
\usepackage{amssymb}
\usepackage{graphicx}
\usepackage{amsmath}
\usepackage[usenames, dvipsnames]{color}
\usepackage{bm}

\begin{document}

\title{Localization length versus level repulsion in 1-D driven disordered quantum wires}

\author{Enrique Benito-Mat\'{\i}as and Rafael A. Molina}

\affiliation{Instituto de Estructura de la Materia, IEM-CSIC, Serrano 123, Madrid 28006, Spain}

%\maketitle
\begin{abstract}
We study the level repulsion and its relationship with the localization length in a disordered one-dimensional quantum wire excited with monochromatic linearly polarized light and described by the Anderson-Floquet model. In the high frequency regime, the linear scaling between the localization length divided by the length of the system and the spectral repulsion is the same as in the one-dimensional Anderson model without driving, although both quantities depend on the parameters of the external field. In the low frequency regime the level repulsion depends mainly on the value of the amplitude of the field and the proportionality between level repulsion and localization length is lost. 

%The low frequency results can be interpreted as a breakdown of single parameter scaling.

%We distinguish two regimes: In the high-frequency regime the ratio between the spectral repulsion and the localization length is the same as in the one-dimensional Anderson model without driving while in the low-frequency regime the level repulsion depends mainly on the value of the amplitude of the field and the results show a break down of single parameter scaling. 
%We study the relation between localization length and spectral statistics (measured through the nearest-level repulsion) in a 1-D Anderson model excited with an harmonic
%time-dependent potential. This relation is known for the 1-D Anderson model, but only in a limited range of the physical parameters. The validity of this relation with 
%extreme accuracy for any system size, lead us to the possibility of studying time-dependent systems, whose size is critical for numerical computations to de able. The 
%results we obtained lead us to interesting questions about the meaning of this relation in physical systems as well as possible applications in quantum wires.
\end{abstract}

\maketitle

\section{INTRODUCTION}

Coherent control through time-dependent periodic external fields is an exciting field in condensed matter physics with promising applications in many different areas like
quantum transport \cite{Lehmann02,Platero04,Camalet04,Kohler05,Martinez08,Gu11,Lopez15}, spintronics \cite{Kirilyuk10,Takayoshi14}, control of many-body phases \cite{Eckardt05,Santos09,Ponte15} and topological properties \cite{Oka09,Kitagawa10,Lindner11,GomezLeon13,Narayan15,Gonzalez16}. 
Some of its guiding principles have already been demonstrated experimentally in different setups 
\cite{Keay95,Madison95,Guhr07,Lignier07,Wang13}. A new and exciting development in the field is the existence of
Floquet time crystals where a discrete translational time symmetry is spontaneously broken in systems with strong disorder and interactions \cite{Else16,Zhang17,Choi17}.

One of the seminal works in the field is the discovery of dynamical localization by Dunlap and Kenkre and the high frequency related phenomenon of coherent destruction of tunneling\cite{Grossmann91,Dunlap86}. The hopping between sites in a lattice subject to a time-dependent potential is renormalized by a Bessel function depending on the field amplitude. At the zeros of this Bessel function the bandwidth of the system goes to zero, the group velocity of a wave-packet vanishes and the particle becomes effectively localized. Experimentally, this effect was first observed as a suppression of current at some amplitudes of the AC electric field \cite{Keay95}. Many of the proposed protocols for coherent control in different quantum systems are based on versions of this effect \cite{Creffield07}.
In particular, in the case of a disordered one-dimensional lattice Holthaus {\em et al.} showed that using the interplay of dynamical localization and disorder, it is possible to control the degree of Anderson localization in the system \cite{Holthaus95,Holthaus96}. 

In mesoscopic systems, where many of the applications of coherent control through time-periodic fields are envisioned, disorder determines or influences most physical properties, and studying the combined effects of disorder and the periodic field becomes crucial for real-world applications. In spite of that, not many works treat both effects on the same footing, probably because of the numerical and theoretical difficulties of ensemble averaging in Floquet theory, the main theoretical tool for studying quantum systems with time-periodic Hamiltonians. 

\begin{figure}[h]
\includegraphics[width=9cm, height=9cm]{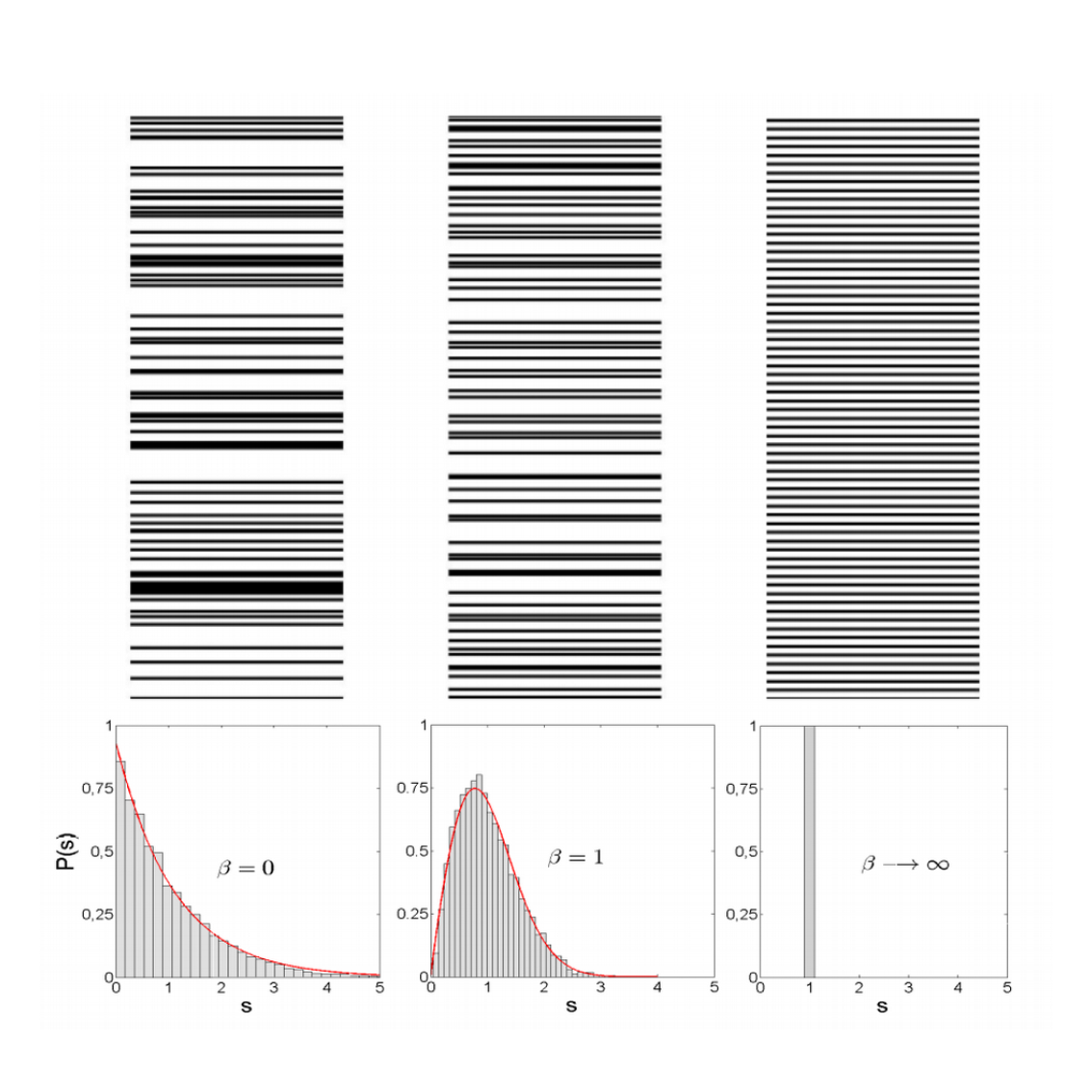}
\caption{\label{fig:espectro} Three different types of behavior for the level repulsion. From left to right: \textit{Poisson distribution} (very large values of disorder with no level repulsion), \textit{Wigner distribution} (intermediate disorder, same as the GOE result), and  \textit{$\delta$-like distribution} (system without disorder) }
\end{figure}

When describing the properties of disordered systems there are two quantities of special interest,
namely, the localization length $l_\infty$ measuring the exponential decay of the wave functions in space,
and the repulsion parameter $\beta$ measuring the repulsion between neighbouring levels in the energy spectrum.
%, and defined as $P(s)\longrightarrow s^\beta$ as $s\longrightarrow0$.
%(See Fig.\ref{fig:espectro})
%$s$ being the energy separation between nearest nein the spectrum and properly normalized so as the mean level spacing be equal to $1$ (see \textbf{Section 3}). 
The localization length is directly related to the conductance in the open system \cite{Abrahams79}, while $\beta$ is usually related to the chaotic behaviour of the classical system \cite{Bohigas84}. For chaotic systems $\beta$ takes the same universal value as the classical random-matrix ensembles depending on the symmetry of the system, $\beta=1$ for orthogonal symmetry, $\beta=2$ for unitary symmetry, and $\beta=4$ for symplectic symmetry \cite{Mehta_Book}. 
However, in disordered lattice models, there is no simple classical analogue but the $\beta$ parameter is still very useful to study localization and can even be used to define the metal-insulator transition \cite{Efetov83,Evers08}.
% defined and in fact, a extremely simple relation with $l_\infty$ is found under some particular conditions. 
In the 1-D Anderson model there is a simple linear relationship between $l_{\infty}/L$ and $\beta$ ($L$ is the total length of the system), implying that both quantities are different measures of the same property \cite{Sorathia12}. The parameter $\beta$ in finite Anderson chains separates from its value in the random-matrix ensembles and takes all possible values between $0$ and $\infty$ depending on disorder and chain length. A pictorial explanation of different kinds of spectra is shown in Fig. \ref{fig:espectro}. This linear relationship has also been found experimentally studying the vibrations of disordered aluminium rods \cite{Flores13}.
%In this case, the qualitative behavior of the $\beta$ parameter as a function of the energy and of the disorder strength can be deduced using some RMT arguments and, in principle, this behavior can be improved using an aproximation for the analytic expression of $l_\infty$ due to Thouless and known as Thouless relation \cite{thouless} (see Apendix A).

In time-periodic models the localization length can be defined using a generalization of the definition for the time-independent case as a single quantity\cite{Martinez06}.
The behavior of the localization length has been studied for the Anderson-Floquet model \cite{Martinez06,Martinez06b}. An increase in the localization length $l_{\infty}$ has been found for low-frequency driving, while $l_{\infty}$ is generally reduced in the high frequency case due to the interplay of disorder and dynamical localization. The implication of these results for the conductance distribution has also been explored \cite{Gopar10,Kitagawa12}. On the other hand, the spectral statistics have not been studied with the same detail, probably because of the difficulties in treating the time-dependent disordered models with large enough sizes to ensure the validity of the usual techniques of unfolding the spectra \cite{Haake_book}.

In this work we fill that void by studying short chains but using an {\it ensemble unfolding}, that is, we calculate the average energy difference between neighboring levels for a given energy as the average over an ensemble of Anderson chains of a certain size. In this way we can make a proper unfolding for small system sizes, as small as $10$ sites long, and take into account the dependence of the spectral statistics on the excitation energy. Focusing on small system sizes, contrary to most works on the topic, we are able to compute spectral statistics for a wide range of frequencies and amplitudes of the driving field, including as many sidebands as needed for good convergence. In the Appendix we show that this kind of unfolding works extremely well for the Anderson model and allows us to recover the linear relationship between $\beta$ and $l_{\infty}/L$ with very good agreement with previous works in the numerical computation of the slope. This paper is organized as follows: In Sec. \ref{sec:anderson-floquet} we introduce the Anderson-Floquet model and discuss the Floquet formalism we use to calculate the localization length. In Sec. \ref{sec:num} we explain the numerical results for different energies and values of disorder distinguishing between the high and low frequency regimes. As an independent estimation of the localization length we also compute the time evolution of a $\delta$-wave packet whose width as a function of time shows periodic oscillations for long times in the low frequency regime.
%
%NEW
%
%We also compute the time evolution of a $\delta$-wave packet as an independent estimation
%of localization lengths.
%
%END
In Sec. \ref{sec:conclusions}, we summarize our results. 
%and present some open questions. 
In the Appendix we apply the same ensemble average method to the Anderson model without driving in order to show that it reproduces extremely well the linear relationship between $l_{\infty}/L$ and $\beta$ even with very small system sizes.

%and it can be proven that 
%Nonetheless, the situation is completely different in time-dependent models. Despite the fact that localization properties of these systems are well known (see for example \cite{dynloc}, \cite{aclocraf}), statistics of the spectra are poorly understood (specifically speaking, the nearest-level repulsion). The essential difficulty deals with numerical computations for typical system size, large enough, to ensure usual techniques of unfolding spectra to be valid for comparison with RMT ensembles. On the one hand, for time-dependent systems it is necessary to deal with very short chains to compute, on the other hand, there is no question about the interest of studying if these quantities are well defined in the same fashion for an
%y system size, most of all, for systems as short as possible. In this work we will study chains with only ten alocations long. With systems as short as ten alocations or less, nothing but an ensemble unfolding makes sense, but the fact this is going to work or 
% not, depends on the 
%density of states. Now, this kind of unfolding works extremely well for the 1-D Anderson model (see Apendix B), and this gave us the motivation to make computations in time-dependent systems where it works as well as expected.

\section{ANDERSON-FLOQUET MODEL}
\label{sec:anderson-floquet}

We describe a one-dimensional quantum wire through a simple single-band tight-binding Hamiltonian. We assume incident monochromatic light coming from a distant source and polarized in the direction of the axis of the wire with a wavelength longer than its length.  
%\textcolor{mypink1}{coming from a distant source.}
%Suppose we have an incident AC field on a 1-D quantum wire with disorder so as the dipolar aproximation of the field to be valid. 
In this framework, the dipolar approximation is valid and we can write a Hamiltonian with the following expression
\begin{align}
H(t)_=\sum_{i=1}^{N}[(\varepsilon_{i}+A i cos(\omega t))\vert i\rangle \langle i\vert\nonumber\\
-J(\vert i\rangle\langle i+1\vert + \vert i+1\rangle\langle i\vert)], \label{eqn:floquethamiltonian}
\end{align}
where $\varepsilon_{i}$ represents the on-site energy for site $i$ which will be considered a random variable uniformly distributed over the range $[-W/2,W/2]$, $J$ is the hopping matrix element between neighboring sites, $A$ is the amplitude of the incident field, and $\omega$ is the frequency of the field
%and $J$ is the "hopping" constant related with tunneling between neighbours 
%and obtained through the expression of the discrete Laplacian 
(throughout this work we use units such that $\hbar=1$, we fix $J=1$ and measure all energies in units of $J$).

Under these conditions we need to solve the complete Schr\"odinger equation:
\begin{align}
H(t)\vert\varphi(t)\rangle=i\hbar \frac{\partial\vert\varphi(t)\rangle}{\partial t}.
\end{align}
For that purpose we can make use of the Floquet theorem that states that for time-periodic Hamiltonians all solutions have the following structure:
\begin{align}
\vert\varphi(t)\rangle=e^{-ie_\alpha t/ \hbar}\vert\varphi_\alpha(t)\rangle,
\end{align}
with $\vert\varphi_\alpha(t)\rangle$ being periodic in time with the same period as $H(t)$, $\vert\varphi_\alpha(t)\rangle=\vert\varphi_\alpha(t+T)\rangle$, $T=2 \pi / \hbar \omega$, and $e_\alpha$ being the quasienergy, a conserved quantity
playing a role similar to the energy in the static Schr\"odinger equation.
We can expand $\vert\varphi_\alpha(t)\rangle$ in Fourier series as
\begin{align}
\vert\varphi_\alpha(t)\rangle=\sum_{n=1}^N\sum_{s=-\infty}^{\infty}C_\alpha^{s,n}\cdot e^{-is\omega t}\cdot \vert n\rangle
\end{align}
Substitution of these terms into the Schr\"odinger equation transforms the equation into an eigenvalue problem for the coefficients $C_\alpha^{s,n}$ and the quasienergy $e_\alpha^m$:
\begin{align}
(\varepsilon_k-m\hslash\omega)C_\alpha^{m,k}-j(C_\alpha^{m,k+1}+C_\alpha^{m,k-1})+\nonumber\\+\dfrac{1}{2}Ak(C_\alpha^{m-1,k}+C_\alpha^{m+1,k})=
e_\alpha^sC_\alpha^{m,k}
\label{eqn:coefficients}
\end{align}
So we can study an \emph{effective Floquet Hamiltonian $H_F$}. In matrix form,
%$H_F = $
\begin{align}
H_F=\begin{bmatrix}
... & ... & ... & ... & ... & ... & ...\\ 
... & (0) & (A) & (\varepsilon + \hslash\omega) & (A)  & ... & ...\\ 
... & ... & (0) & (A) & (\varepsilon) & (A) &...\\ 
... & ... & ... & (0) & (A) & (\varepsilon - \hslash\omega) & ...\\ 
... & ... & ... & ... & ... & ... & ...
\end{bmatrix}\label{eqn:floquethamiltonian2} 
\end{align}
where $(A)=\begin{bmatrix} \dfrac{1}{2}A &0&...&0\\
0&.&...&0\\
.&...&.&.\\
0&...&...&\dfrac{1}{2}NA \end{bmatrix}$ and \\\\\\
$(\varepsilon + n\hslash\omega)=\begin{bmatrix}\varepsilon_1 + n\hslash\omega&-J&...&0\\
-J&\varepsilon_2 + n\hslash\omega&-J&.\\
.&.&.&.\\
0&.&-J&\varepsilon_N + n\hslash\omega\end{bmatrix}$
\\
\\
\\
This matrix $H_F$ is infinite dimensional by virtue of the $m$-label denoting the different modes of the expansion, so we need to truncate this matrix for computations. We discuss the actual criteria of convergence in the next section but we have checked that the quantity of interest in the calculations (the repulsion parameter $\beta$, for example) was stable as the number of modes included in the Hamiltonian was increased. 

%As the spectrum is periodic with period $\omega$ we use the equality of the spectra of the adjacent copies of the central band as a criteria for deciding how many modes we need to ensure convergence. We
%take as many modes as needed for the spectra of the bands $m=1,-1$ is equal to the spectra of the first Floquet-Brillouin zone adding $m \hbar \omega$ (within some small tolerance of $10^{-8}$).
%%% WHAT IS THE TOLERANCE IN PRACTICE

%(Although there are some properties we can understand with this picture, the system is not equivalent to a 2-D system in all regimes as we will see).

%The resulting spectrum is made of equal energy bands corresponding to different modes with a "band gap" equal to the energy of the incident photons $\hslash\omega$. Taking this into account we can restrict ourselves to the study of the equivalent "First Brillouin zone" as that part of the spectrum that belongs to the zero-mode. 
%Then, we have two distinct and clearcut regimes depending on the energy of the incident photon is enough, or not, to take one of the eigenvalues outside the support of the energy band. Throughout this work, we will study both regimes \emph{High frequency regime} ($\omega>(w+4j)/2$) and \emph{Low frequency regime} ($\omega<(w+4j)/2$)

Without loss of generality we can restrict ourselves to the calculation of the level repulsion and the localization length of the states in the first Floquet-Brillouin zone. There are then two distinct regimes depending on whether or not the energy of a photon of the external field is enough to take all the states outside the energy band of the undriven system. In the former case there is an energy gap between the different Floquet-Brillouin bands while there is no gap in the later. Throughout this work we study both, the \emph{high frequency regime} ($\omega>(W+4J)/2$) and the \emph{low frequency regime} ($\omega<(W+4J)/2$).
% (see Fig. \ref{fig:fig10}).

% Probablemente haya que poner fig10 aqui

%\begin{figure}[!htb]
%	\includegraphics[width=1\columnwidth, height=4.5cm ]{spectrum}
%	\caption{\label{fig:fig10} Spectral results in the \textit{high frequency regime} and the \textit{low frequency regime} that shows how the "band gap" dissapears at low freq. causing levels at different bands to "interact" with each other and thus, changing the behavior of $\beta$ parameter as compared with the \textit{high frequency} case.}
%\end{figure}	

\section{NUMERICAL RESULTS}
\label{sec:num}
%We will make computations with three different disorders $w={2,5,10}$ ($\hslash$ units) and, at least, with four different values of the adimensional parameter $\dfrac{A}{\omega}=0, 0.004, 0.04, 0.4$ in a chain with ten alocations. Low frequency results need more calculations in order to describe properly its behavior, as we will see.
\subsection{Computation of level repulsion}
%Except for the classical Gaussian ensembles and some other extensions (the pioneering papers by Wigner, Dyson, Porter, Gaudin and others are collected in \cite{porter}, extensions to other ensembles are reviewed in \cite{brody}) there are no analytical results for the $\beta$ parameter in general.
In order to compute the level repulsion we diagonalize the Floquet Hamiltonian (\ref{eqn:floquethamiltonian2}). The number of modes needed for convergence depends on system size, frequency and amplitude of the field. We have used a very strict criterion for convergence. The relative error in the spacings between the results in the central Floquet band and the two adjacent ones is smaller than $10^{-4}$ for all spacings. 
%is chosen such that two adjacent Floquet bands to the central one, have the expected $\pm\hslash\omega$ difference in the quasienergies within a standard error $<10^{-4}:\Delta e\mp\hslash\omega=e_{\pm1}^m-e_0^m\mp\hslash\omega<10^{-4}$}. 
In our model we find empirically that $N_{modes} \sim k \dfrac{AL}{\omega}$ \cite{Martinez06} with $2< k <3$.  For example, the number of modes we have used for $A=4$ $\omega=1$ and $L=10$ is $103$ ($k \simeq2.575$), which amounts to the diagonalization of matrices of dimension $d=1030$.

\begin{figure}
\includegraphics[width=1.05\columnwidth, height=5.5cm]{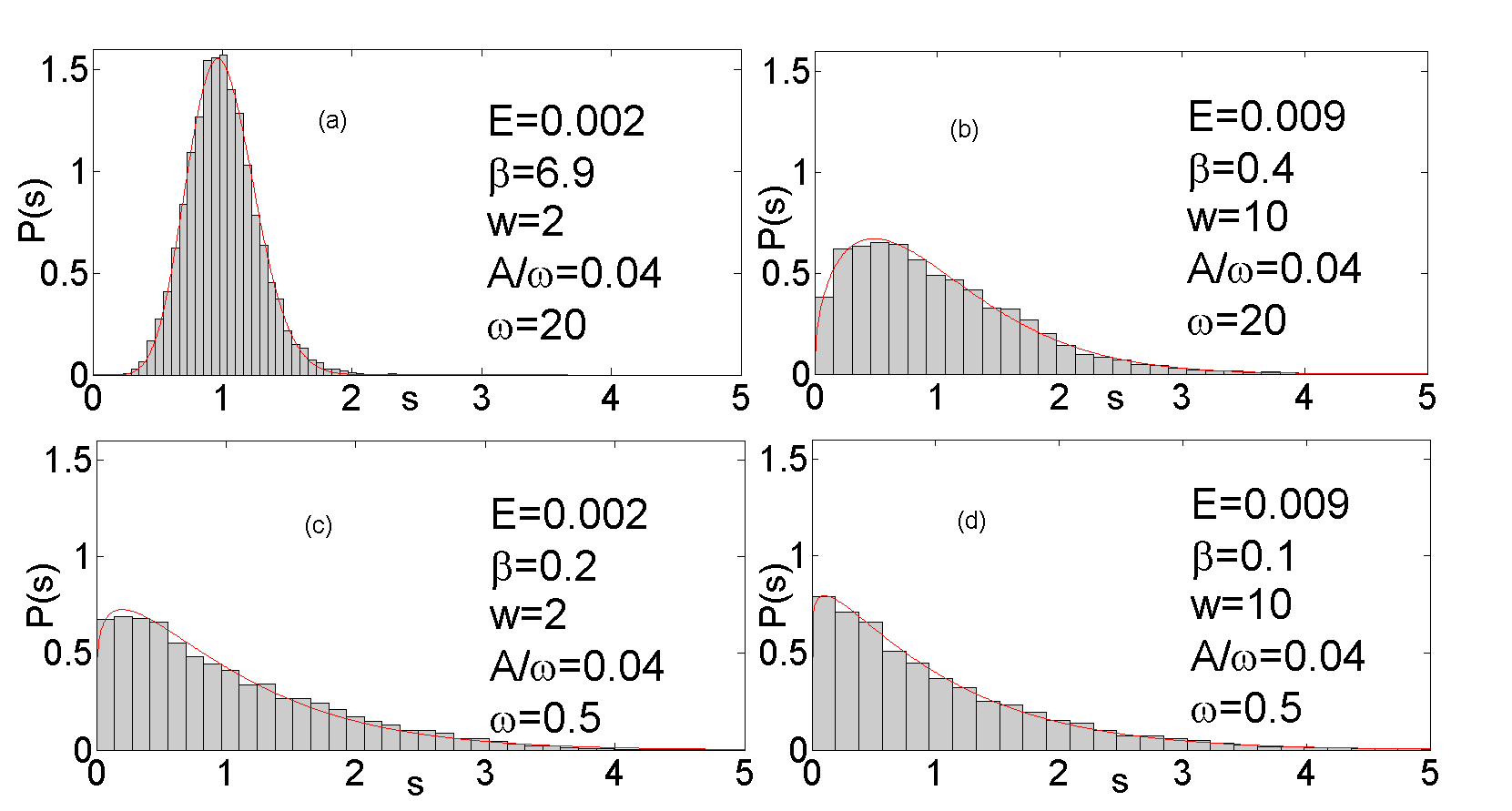}
\caption{\label{fig2:fig2} Nearest-neighbor spacing distributions $P(s)$ (histograms) and fits to Eq. (\ref{eq:Pofs}) (red solid lines) for energies $E$ close to the band center, two different values of the disorder strength $W$ and a fixed value of $\frac{A}{\omega}$:(a) and (b) high frequency examples ,and (c) and (d) low frequency examples.}
\end{figure}

To construct the nearest-neighbor spacing distribution $P(s)$ we include the results for 10000 realizations for each value of the disorder strength, frequency and amplitude of the field. 
%Although there are better distributions to compare with, \cite{izra1} we will use two different models, depending on the $\beta$ values. 
%For $\beta>1$ we will use a \emph{Wigner distribution}
%$P(s)=as^\beta e^{-bs^2}$
%and for $\beta<1$ we will use a \emph{Brody distribution}
%$P(s)=c(1+\omega)s^\omega e^{-cs^{(1+\omega)}}$
%The reason is that these two models fit extremely well with our numerical data and are much more simple than that in \cite{izra1}. Besides this, it should be remembered that we are interested in the behavior of $P(s)$ as $s\longrightarrow0$ so, in principle, no more information than $\beta$ parameter should be obtained from these comparisons.
%P(S) PARA W=2,10 A/omega=0.04 en alta(arriba) y baja frecuencia(abajo) pARA E aprox. =0
We are most interested in the repulsion parameter $\beta$ defined through the behavior of $P(s)$ for small spacings $P(s) \sim_{s \rightarrow 0} s^{\beta}$. However, we obtain a very good estimation of $\beta$ through fits of the whole distribution to the following model:

\begin{equation}
P(s)=\begin{cases}
   c(1+\beta)s^\beta e^{-cs^{(1+\beta)}}, & 0< \beta <1, \\
   as^\beta e^{-bs^2}, & \beta>1,
%\text{otherwise},
   \end{cases}
\label{eq:Pofs}
\end{equation}
where $c$, $a$, and $b$ are $\beta$-dependent normalization constants that are obtained through the conditions $\int ds P(s)=1$ and $\int ds s P(s)=1$. For the classical Random Matrix Ensembles, GOE (Gaussian Orthogonal Ensemble), GUE (Gaussian Unitary Ensemble), and GSE (Gaussian Symplectic Ensemble)$\beta=1$, $2$, and $4$ respectively while in the Poisson case $\beta=0$ \cite{Mehta_Book, Brody81,Porter65}. As previously mentioned for the finite-size Anderson model, however, $\beta$ can take all values from $\beta=0$ for very large values of disorder to $\beta=\infty$ in the clean system.  The continuous Gaussian ensemble or Hermite Ensemble has been proposed for describing the properties of systems with continuous values of the level repulsion parameter with a similar description of the $P(s)$ \cite{Dimitriu02,Dimitriu05,Relano08}.  This proposed formula produces high-quality fits for the whole range of spacings as we show in some examples below. The part for $\beta< 1$ is the widely used Brody distribution, which is normally used to fit the spacing distribution in
the transition between integrability and chaos \cite{Brody73}. The part for $\beta>1$ is a generalization of the Wigner distribution usually applied for comparison with the classical Random Matrix Ensembles. Other distributions can be used that fit the whole range of values for the repulsion parameter, for example, the Izrailev distribution \cite{Izrailev88,Izrailev89}. For our model they produce very close results with fits of similar quality but are more complex to use as the normalization constants do not have a closed analytical form. In Fig. \ref{fig2:fig2}, several representative examples of these fits are shown in the low and high frequency regimes. The high quality of the fits is remarkable even if the fitting function is only phenomenological and gives further justification for the use of the ensemble unfolding. 
 
\subsection{Computation of the localization length}

For the computation of $l_\infty$ we use a Floquet-Green function formalism developed by D. F. Martinez \cite{Martinez03}. A Floquet-Green operator corresponding to (\ref{eqn:coefficients}) can be defined as

\begin{align}
(E+m\hslash\omega-(\varepsilon))G_{1L}^{m}-\hat{V}(G_{1L}^{m-1}+G_{1L}^{m+1})=
\delta_{m,0}
\label{eqn:green}
\end{align}

The different quantities $G_{1L}^m$ are associated with the probability of a process where an electron starts with an energy $E$ at site $1$ and ends at site $L$ with energy $E+m\hslash\omega$.
The transport properties of driven systems have been formulated in terms of this Floquet-Green operator, which plays a role similar to the Green's function in the Landauer formalism for conduction \cite{Kohler05}.

The definition of localization length in terms of the Green's function \cite{Kramer93} can be generalized for the Floquet-Green function operator \cite{Martinez06,Martinez06b}, where the double overline means the average over different realizations:
But it can be shown that in the thermodynamic limit ($L \to +\infty$) the values of $l_{\infty}^{m}(E)$ for all the different values of $m$ are the same. That is, localization lengths become independent of $m$ (see \cite{Martinez06}) so from now on, we focus our attention on $l_{\infty}^0(E)$, and we rename it $l_{\infty}(E)$ for the sake of simplicity.

In solving (\ref{eqn:green}) for $G_{1L}^0 (E)$ we can use matrix continued fractions \cite{Pastawski83,Pastawski01,Martinez03}. In this case, we get
\begin{align}
G_{1L}^0 (E)=[E-(\varepsilon)-\hat{V}_{eff}(E)]^{-1}
\end{align}
with
\begin{align}
\hat{V}_{eff}(E)=\hat{V}_{eff}(E)^+ + \hat{V}_{eff}(E)^-
\end{align}
and
\begin{equation}
\begin{aligned}
\hat{V}_{eff}(E)^{\pm}=\\
=\hat{V}\dfrac{1}{E\pm \hslash \omega - (\varepsilon) - \hat{V}\dfrac{1}{E\pm 2 \hslash \omega - (\varepsilon) - \hat{V}\dfrac{1}{...}\hat{V}}\hat{V}}\hat{V}
\label{eqn:contfrac}
\end{aligned}
\end{equation}
As with the computation of level repulsion, the number of modes needed to ensure convergence of (\ref{eqn:contfrac}) increases linearly with the parameter $\dfrac{AL}{\omega}$ 
%\textcolor{mypink1}{($N_{modes} \sim k_{l_{\infty}}
%\dfrac{AL}{\omega}$)} 
and was explored in previous works \cite{Martinez06,Martinez06b}. %\textcolor{mypink1}{In this sense, it is found that the proportionality constant needed for convergence in this case ($k_{l_{\infty}}$) is such that $k_{l_{\infty}}\simeq k_\beta/4$. (This is: convergence for the Green's function is different than convergence for the eigenvalues) }
% but, in contrast with $beta$ parameter, there are either analytical or semi-analytical results for $l_\infty(E)$  under different physical conditions \cite{Thouless79,Holthaus95}.
\begin{figure}
\includegraphics[width=0.95\columnwidth]{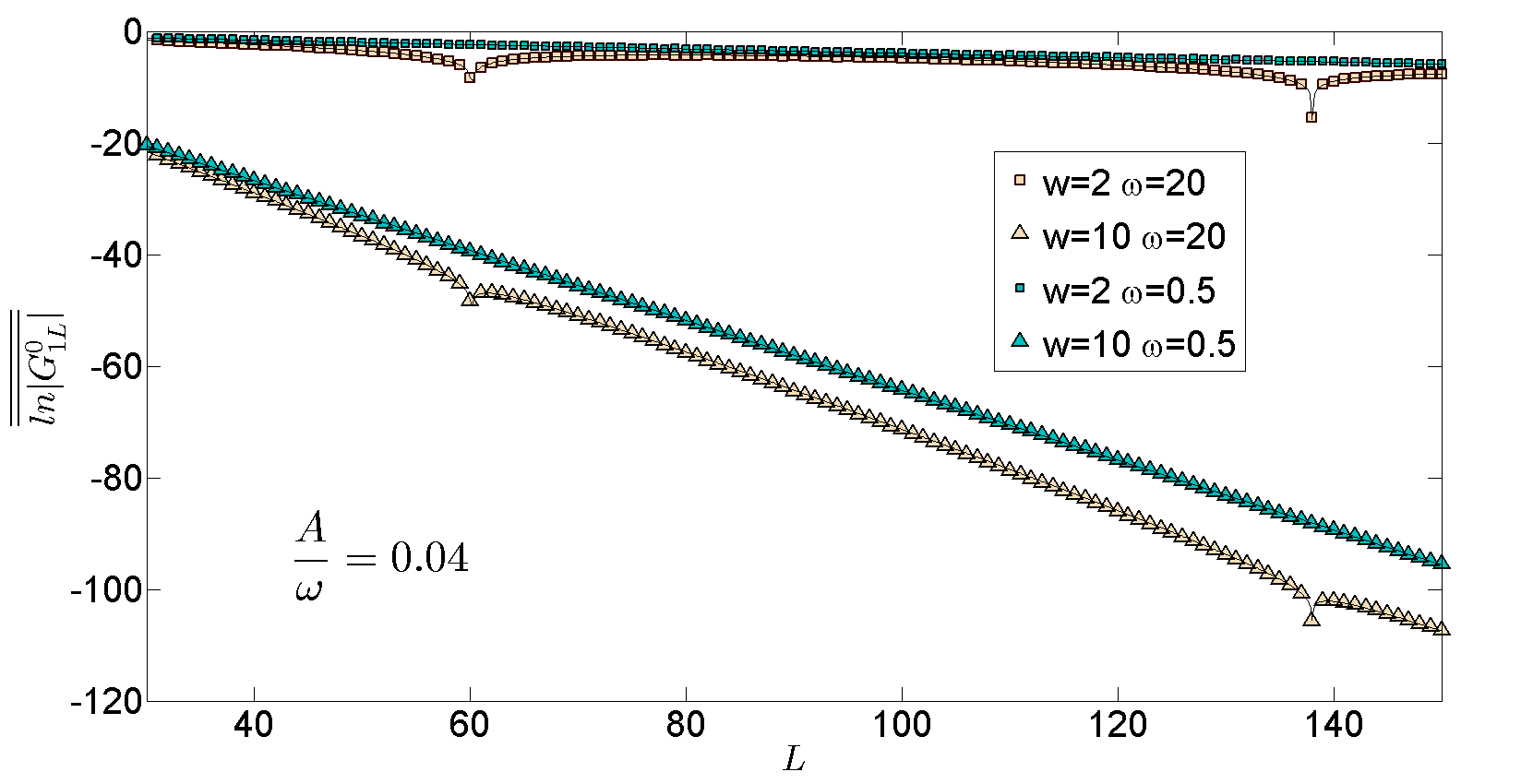}
\caption{\label{fig:fig3} Ensemble average $\overline{\overline{ln\vert G_{1L}^0 (E)\vert}}$ vs $L$ for the calculation of the localization length $l_\infty$ for the same parameters as in Fig. \ref{fig2:fig2}. Here the value of $l_\infty$ is the inverse slope in this figure.}
\end{figure}
Examples of the numerical results involved in computing $l_\infty(E)$ are shown in Fig. \ref{fig:fig3}. In the low frequency cases there is an exponential decrease of the value of the Green's function $G_{1L}^0(E)$ signature of Anderson localization. From a fit of the slope in log scale we extract the value of the localization length. In the high frequency cases the exponential decrease is modulated by a Bessel function due to the dynamical localization and the fit has to been done accordingly \cite{Martinez06}. The position of the zeros of the Bessel function is clearly seen in Fig. \ref{fig:fig3} 

\subsection{Temporal evolution of an impulse wave-packet in the Anderson-Floquet model}

We have also calculated the temporal evolution of a $\delta$-impulse through the system. These calculations serve two different purposes: they are a consistency test of the results obtained using $G_{1L}^0 (E)$ for computing $l_\infty(E)$, and they present new results by themselves as in the low frequency regime we observe oscillations that persist in the asymptotic limit. As the Hamiltonian (\ref{eqn:floquethamiltonian}) is such that it does not commute for different times ($[H(t_2), H(t_1)]\neq 0$) we compute the evolution operator $U(t,0)$ numerically. We evolve an initial $\delta$-impulse positioned at time $t=0$ at the middle site of the system. In order to estimate $l_\infty$ from time evolution, we represent the average over 1000 realizations of the mean-square displacement $\left(\overline{\overline{\Delta x(t)}}=\overline{\overline{\sqrt{\langle x^2\rangle-\langle x^2\rangle}}}\right)$ as a function of time. Since all eigenstates of the Hamiltonian (\ref{eqn:floquethamiltonian}) are exponentially localized, we expect an absence of diffusion for long times when $\Delta x \sim l_\infty$. However, the initial state has components at different energies, so the localization length that is probed by these simulations is a mixture of the actual energy-dependent localization length $l_\infty(E)$. For a discussion concerning how to relate localization length to the mean-square displacement in time-independent one-dimensional systems, see for example, \cite{Izrailev97}.

\subsection{High frequency regime}
The localization length has been well studied in the high frequency regime $\cite{Holthaus95,Holthaus96,Martinez06,Martinez06b}$. The Floquet-Anderson model in this regime is known to behave as a 1-D Anderson model with a renormalized hopping constant that shuts off the tunneling between the sites $i-1$ and $i$ at the zeros of the Bessel function $J_0(\frac{A}{\omega})$, a manifestation of coherent destruction of tunneling. In particular, $l_\infty(E)$ can be obtained from the results of the Anderson model without driving by increasing the disorder by this factor.

\begin{figure}[!htb]
	\includegraphics[width=0.95\columnwidth]{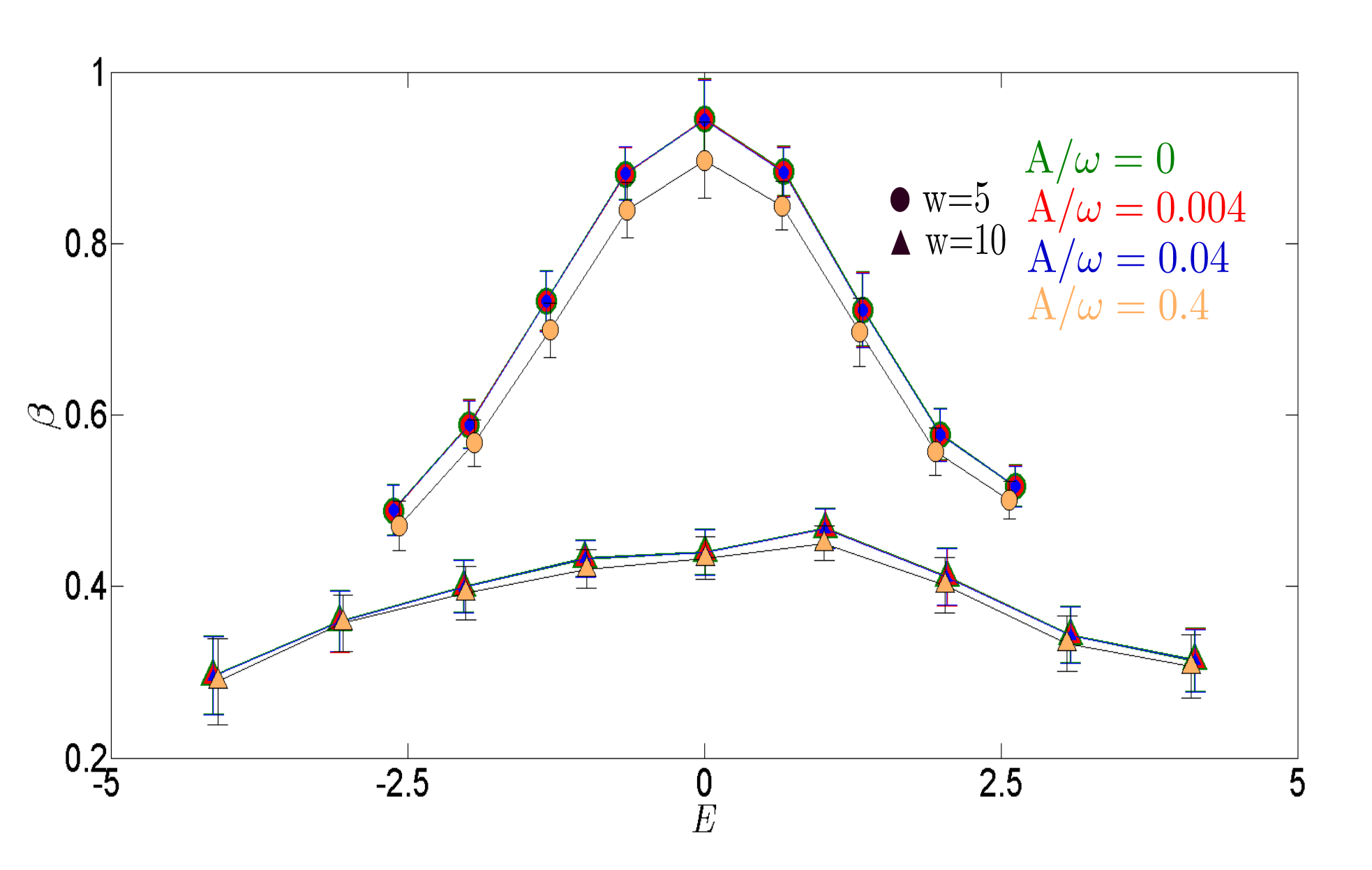}
        \includegraphics[width=0.95\columnwidth]{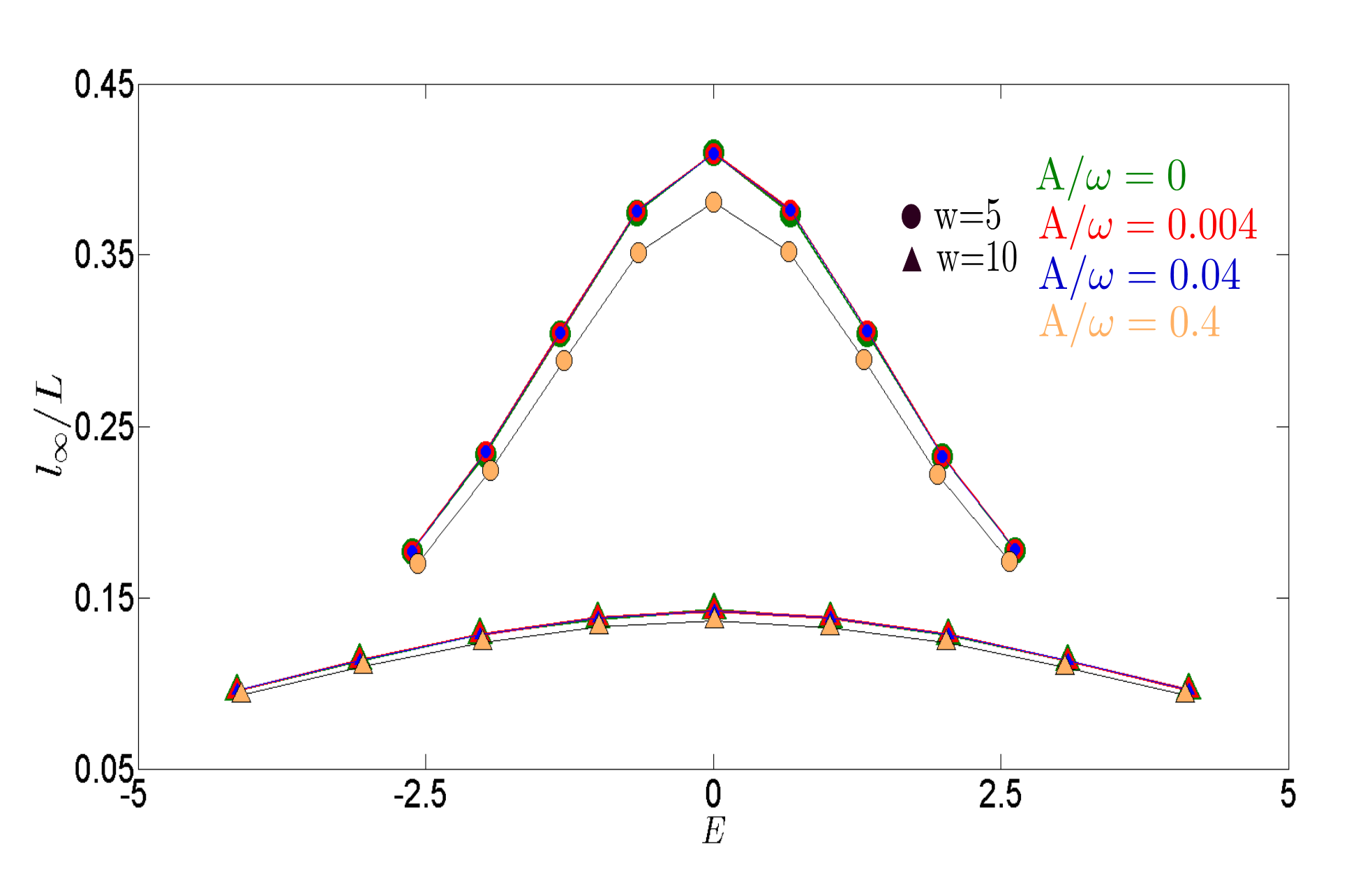}
	\caption{\label{fig:fig4} \textit{High frequency regime} for different values of the adimensional parameter $\dfrac{A}{\omega}$. Top panel: $\beta$ vs Energy. Bottom panel: $\dfrac{l_\infty}{L}$ vs Energy.  }
\end{figure}

In the high frequency regime, the repulsion parameter $\beta$ has the same dependence on the parameters of the model as $l_\infty$, as shown in Fig. \ref{fig:fig4}. %on the physical pin the high frequency regime% 
%\begin{figure}[!htb]
%	\includegraphics[width=1\columnwidth, height=5.8cm ]{lambdavsEhigh105v4}
%	\caption{\label{fig:fig5}$\dfrac{l_\infty}{L}$ vs Energy in the \textit{high frequency regime} for different values of the adimensional parameter $\dfrac{A}{\omega}$}
%\end{figure}
\begin{figure}[!htb]
	\includegraphics[width=1.1\columnwidth]{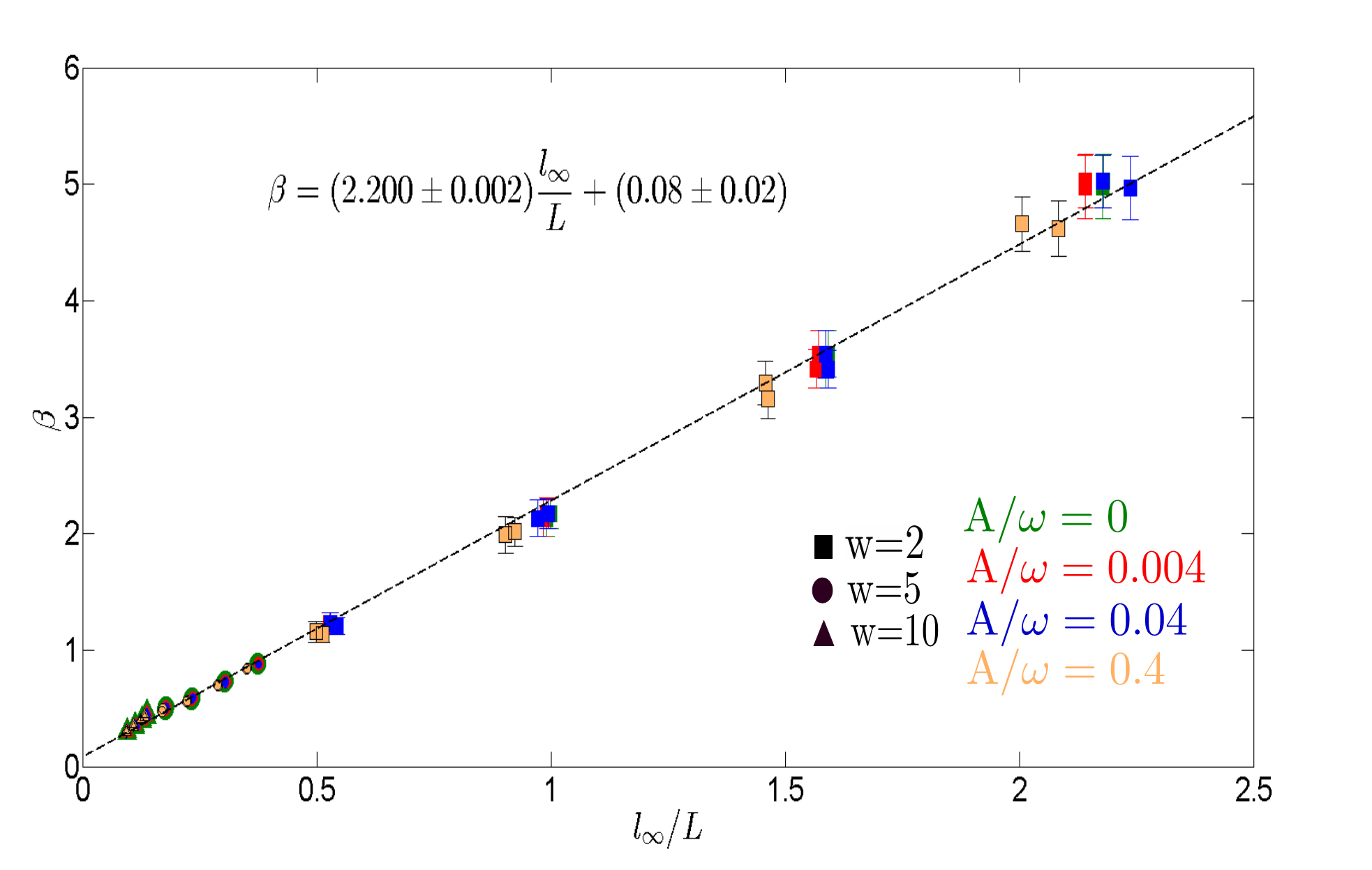}
	\caption{\label{fig:fig6} $\beta-l_\infty$ relation in the \textit{high frequency regime} for three different values of the disorder strength $W$ and four different values of the adimensional parameter $\frac{A}{\omega}$}
\end{figure}

\begin{figure}[!htb]
	\includegraphics[width=0.95\columnwidth]{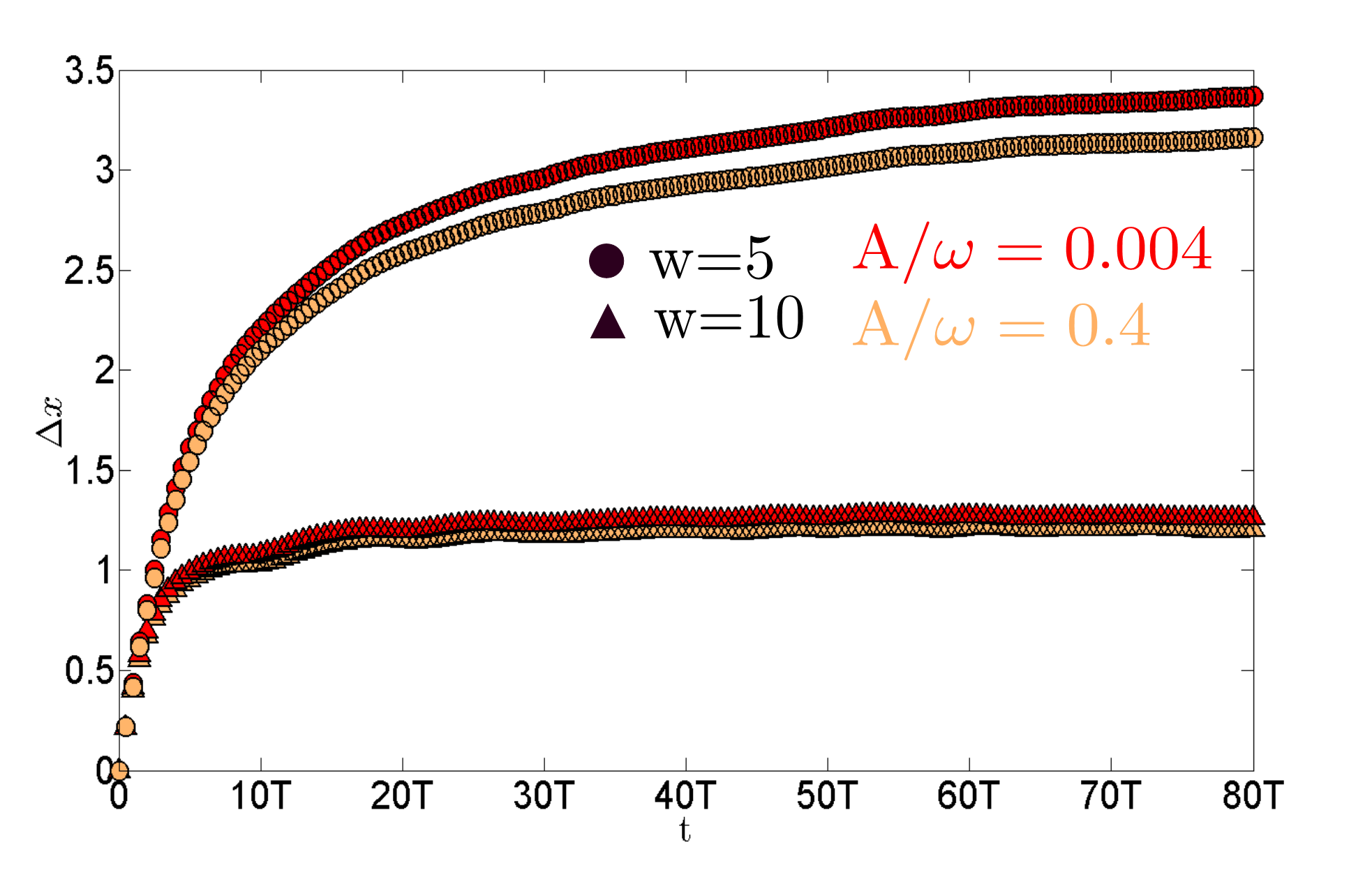}
	\caption{\label{fig:fig5} \textit{High frequency regime}: $\Delta x$ as a function of time measured in periods of the external field $T$ for different values of the disorder strength and the adimensional parameter $\frac{A}{\omega}$.}
\end{figure}

As a consequence, there is a simple linear relationship between the $\beta$ parameter and $\frac{l_\infty}{L}$, (Fig. \ref{fig:fig6}). Moreover, this is (up to small numerical errors) the same relationship as in Anderson's hopping model (see \cite{Sorathia12} and the Appendix). In essence, there is a renormalization of the hopping
by the Bessel function $J_0(\frac{A}{\omega})$ which amounts to a renormalization of the effective disorder strength $W_{eff}=W/J_0(\frac{A}{\omega})$ that affects equally the localization length and the spectral statistics. 

There is some subtle question regarding the $E\simeq0$ states that we should mention now. It is well known that close to the band edges and band center of the Anderson model \emph{Single Parameter Scaling} no longer holds and this leads to anomalous localization values \cite{Altshuler03,Thouless79}. In our numerical results there is a small difference in the slope of the $\beta-l_\infty/L$ relation for these zero-energy states so they are not shown in Fig. \ref{fig:fig6}. In this sense, we found a relation $\beta\simeq2.3\dfrac{l_\infty}{L}$ for zero energy states, instead of $\beta\simeq2.2\dfrac{l_\infty}{L}$.Determining whether this small difference is caused by the anomalous behavior near $E\simeq0$ or not would require further analysis and is not essential for our discussion.

%NEW:

Fig. \ref{fig:fig5} shows the time evolution of a $\delta$-wavepacket for different representative cases in the high frequency regime. The values obtained for $\Delta x(80T)$ are in very good agreement with the calculations of $l_\infty(E)$ using the Floquet-Green function method. The values for the mean-square displacement reach a value close to the maximum for the localization length corresponding to $E=0$. However, because of the energy dependence in $l_\infty(E)$ the limiting value of $\Delta x$ represents an average value for $l_\infty(E)$ over the different energy eigenstates that form the initial $\delta$-wave packet.

%END

\subsection{Low frequency regime}

\begin{figure}[!htb]
	\includegraphics[width=0.95\columnwidth]{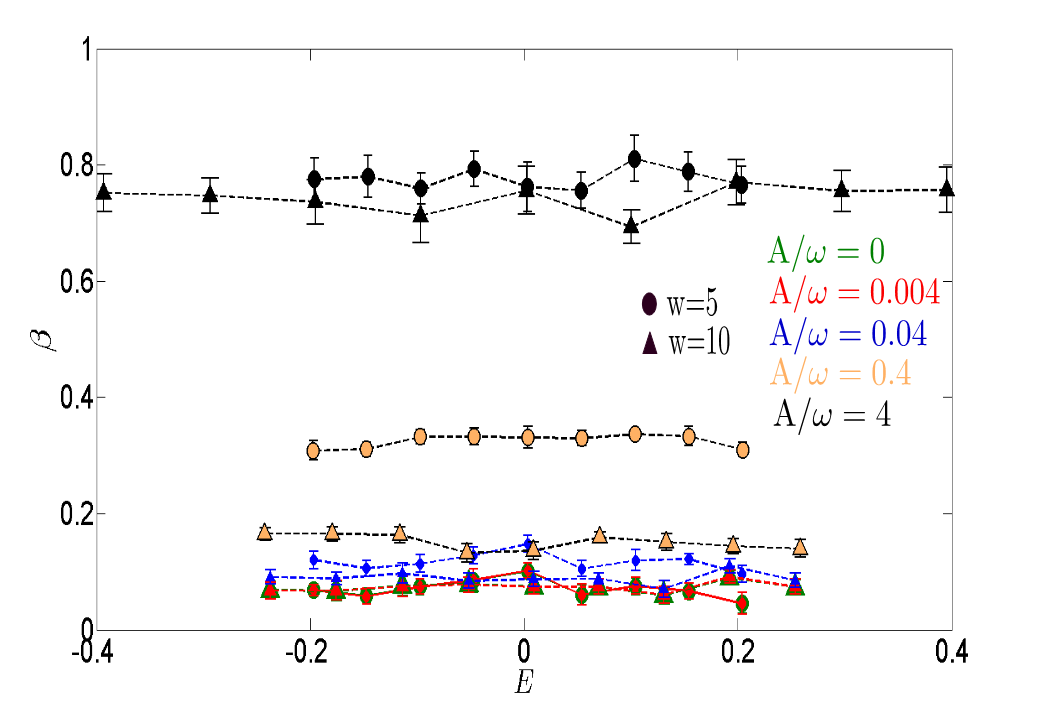}
        \includegraphics[width=0.95\columnwidth]{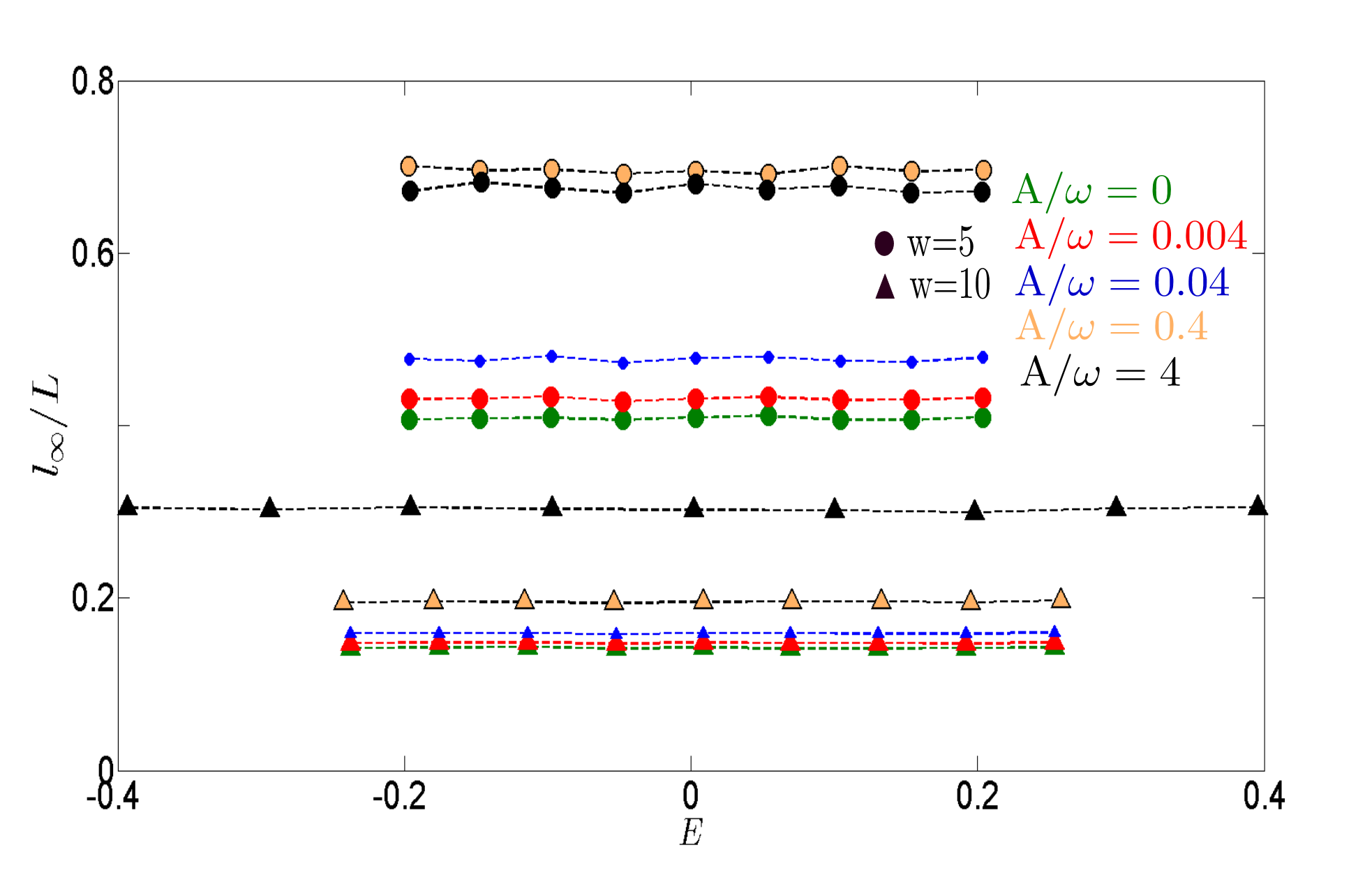}
	\caption{\label{fig:fig7} \textit{Low frequency regime} for different values of the adimensional parameter $\frac{A}{\omega}$ ($\omega=0.5$). Top: $\beta$ vs. energy. Bottom: $l_\infty$ vs energy.}
\end{figure}
\begin{figure}[!htb]
	\includegraphics[width=0.95\columnwidth]{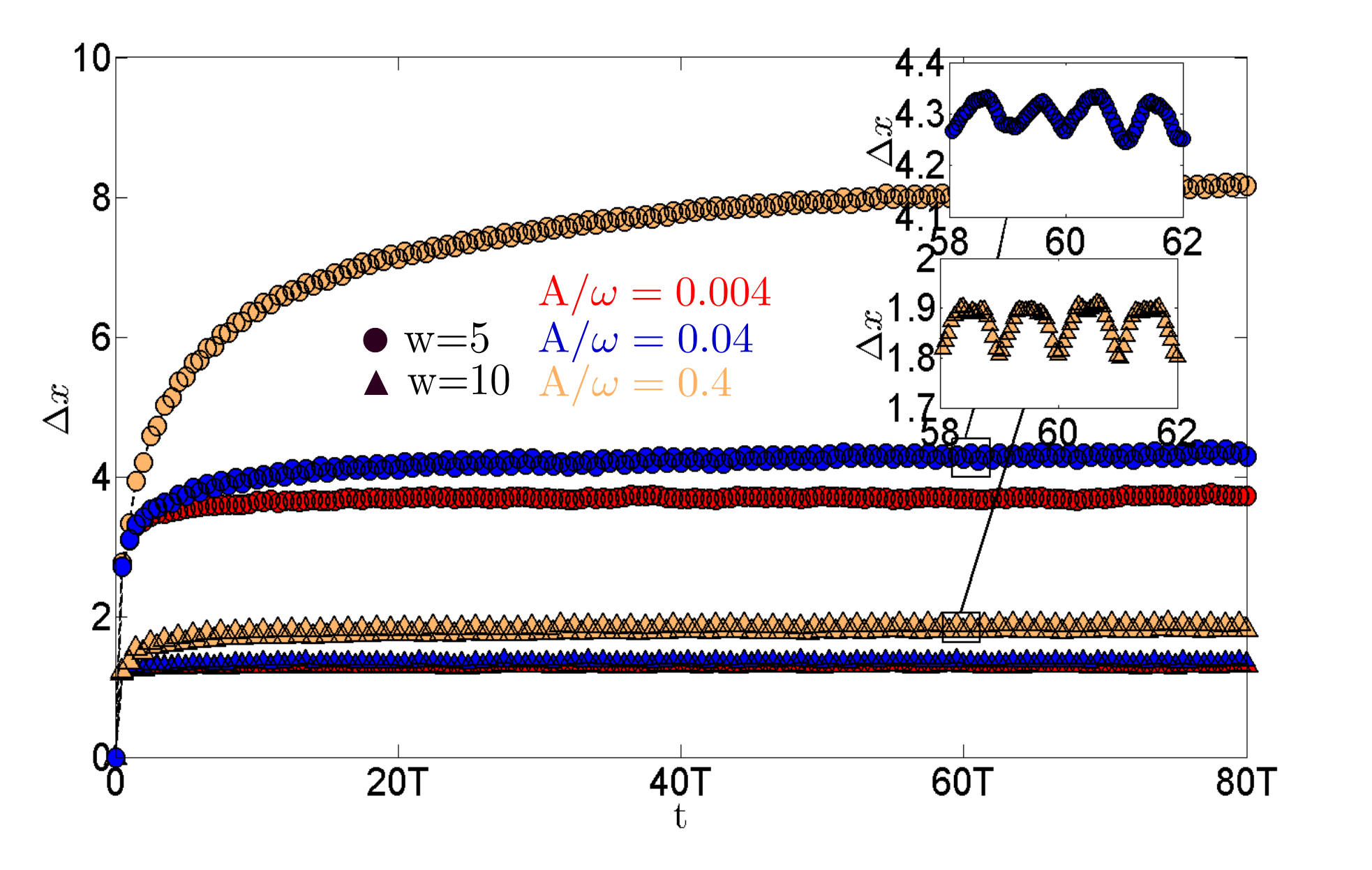}
	\caption{\label{fig:fig8} \textit{Low frequency regime}: $\Delta x$ as a function of time measured in periods $T$ of the external field for different values of the disorder strength $W$ and of the adimensional parameter $\frac{A}{\omega}$. The insets show a zoom of the oscillations within a period for some of the cases considered.}
\end{figure}

%\begin{figure}[!htb]
%	\includegraphics[width=1\columnwidth]{oscilacion}
%	\caption{\label{fig:oscilacion} \textit{Low frequency regime}: $\Delta x$ as a function of time and evolution of the wave packet inside a period for the inset of Fig. \ref{fig:fig8}  corresponding to $W=5$. In the insets the profile of the field through the system along different times is shown together with the wave packet. As hopping constant is $J=1$, the diffusion time $\sim 1/J$ and the period of the field $T=2\pi/\omega$ are of the same order. Consequently, the wave packet evolves in a quasi-adiabatic regime in wich the effective size of the system is periodically diminished by the incident field. This causes the wave packet to spread out or shrink depending on the time within the period.}
%\end{figure}
%\begin{figure}[!htb]
%	\includegraphics[width=0.95\columnwidth]{lambdavsElow105v4}
%	\caption{\label{fig:fig8}$\l_\infty$ vs. energy in the low frequency regime for different values of the adimensional parameter $\dfrac{A}{\omega}$}
%\end{figure}

The study of the low frequency regime is the most challenging since analytical results do not exist and the numerical convergence as a function of the number of modes is slower. Nonetheless, low frequency behavior is richer, as we show in this section. 

In Fig. \ref{fig:fig7} we plot $\beta$ and $l_\infty$ as a function of the energy for different values of $A/\omega$ and disorder strength $W$. An analysis of the examples shown in the figures reveals several features of the low frequency regime. The values of $\beta$ and $l_\infty$ become nearly Energy-independent \cite{Martinez06b}. Low frequency fields tend to delocalize (increase $l_\infty$) compared to high frequency fields, at least for not too high values of the field amplitude $A$. This is in perfect agreement with previous results \cite{Martinez06}. In contrast, disorder has little influence on $\beta$ compared with the influence of the external field amplitude $A$, especially, for high amplitudes. So, in determining level repulsion, the incident field is the dominant term, compared to disorder.

%NEW:

We have also calculated the time evolution of the $\delta$-wavepacket in the low frequency regime. Results for some representative cases are shown in Fig. \ref{fig:fig8}. As for the high frequency regime, the results obtained for $\Delta x(80T)$ are in very good agreement with those obtained for $l_\infty(E)$. In this case, however, the time evolution is more complex than that in the high frequency case. Above a certain value of the field amplitude $A\gtrsim 0.02$ the low frequency evolution exhibits periodic oscillations of the mean-square displacement with period $T=\dfrac{2\pi}{\omega}$. The amplitude of these oscillations is nearly independent of the field strength but small enough that it does not affect the estimation of the localization length from the temporal evolution. %These oscillations are not a finite size-effect of the system and the amplitude is nearly independent of the field but small enough so, localization length can still be estimated from temporal evolution for all cases of interest. 
The origin of such oscillations can be understood as an effect of the matching of two characteristic times, the hopping time related to $1/J$ and the characteristic period of the field $T=2\pi/\omega$. At low frequencies when the period is larger than the typical hopping time, the wave packet undergoes a quasi-adiabatic evolution that causes periodic reflections of the wave packet that decrease the mean-square displacement. 
%Since the field enters in the Hamiltonian (\ref{eqn:floquethamiltonian}) through a time-dependent potential that is proportional to the site inside the wire, these oscillations persits in the thermodynamic limit $L\rightarrow \infty$.

% END
 
\begin{figure}[!htb]
	\includegraphics[width=1\columnwidth]{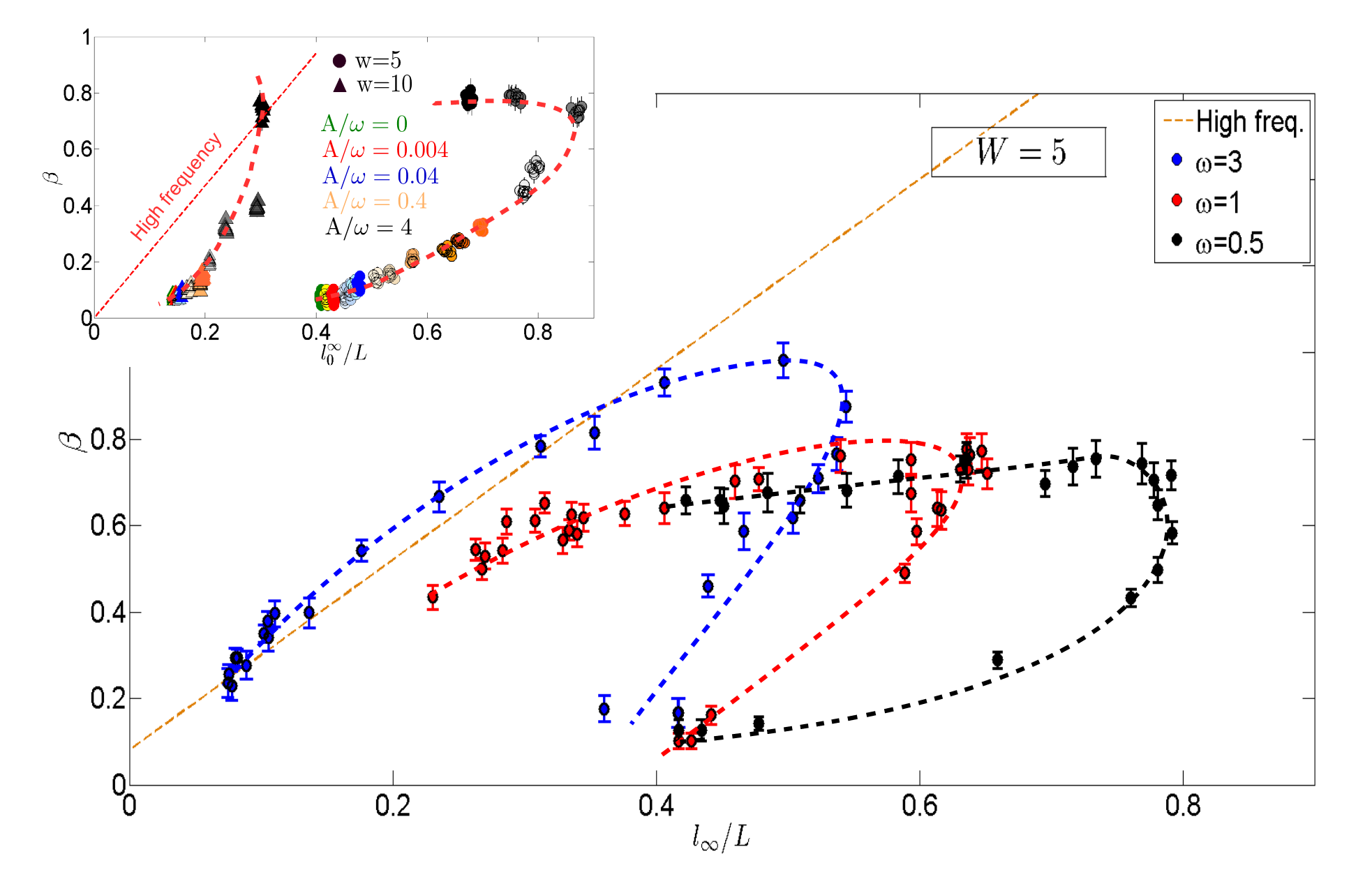}
	\caption{\label{fig:fig9} The $\beta-l_\infty$ relation in the \textit{low frequency regime} for disorder strength $W=5$ and different values of the adimensional parameter $\frac{A}{\omega}$. The linear relationship in the high frequency regime is shown as an orange line. The inset shows the results for two values of the disorder strength $W=5,10$ and only one value of the frequency $\omega=0.5$. Different values of the energy are mixed in the plot.}
\end{figure}

 At low frequency, the simple linear relation between $\beta$ and $l_\infty$ does not apply any more due to the different influences of disorder and amplitude of the field in localization and level repulsion. In Fig. \ref{fig:fig9} we show the relationship between $\beta$ and the ratio $l_\infty/L$ in the low frequency regime. In the main panel we show how this relationship evolves as we reduce frequency for disorder strength $W=5$, while in the inset we show the relationship between $\beta$ and $l_\infty$ for two different values of the disorder parameter, $W=5, 10$, at the same frequency $\omega=0.5$. In both panels the linear relationship for the high-frequency case is also shown for comparison. The points are joined by a guide to the eye, and increasing amplitudes follow a counterclockwise direction in this line. For $\omega=1$ and $\omega=0.5$ the values of $\beta$ increase with increasing amplitude until they are close to $\beta=0.8$ for $A/\omega=4$. The value $\omega=3$ is close to the high frequency limit, and the results are closer to the high-frequency line. The localization length increases with amplitude until $A/\omega \approx 1$ and then decreases again for higher field amplitudes \cite{Martinez06}. The resulting curves present higher curvature for smaller values of disorder strength.  

%The linear $\beta$-$l_\infty$ relation to break down (Fig. \ref{fig:fig9})}.

\begin{figure}[!htb]
	\includegraphics[width=1.1\columnwidth]{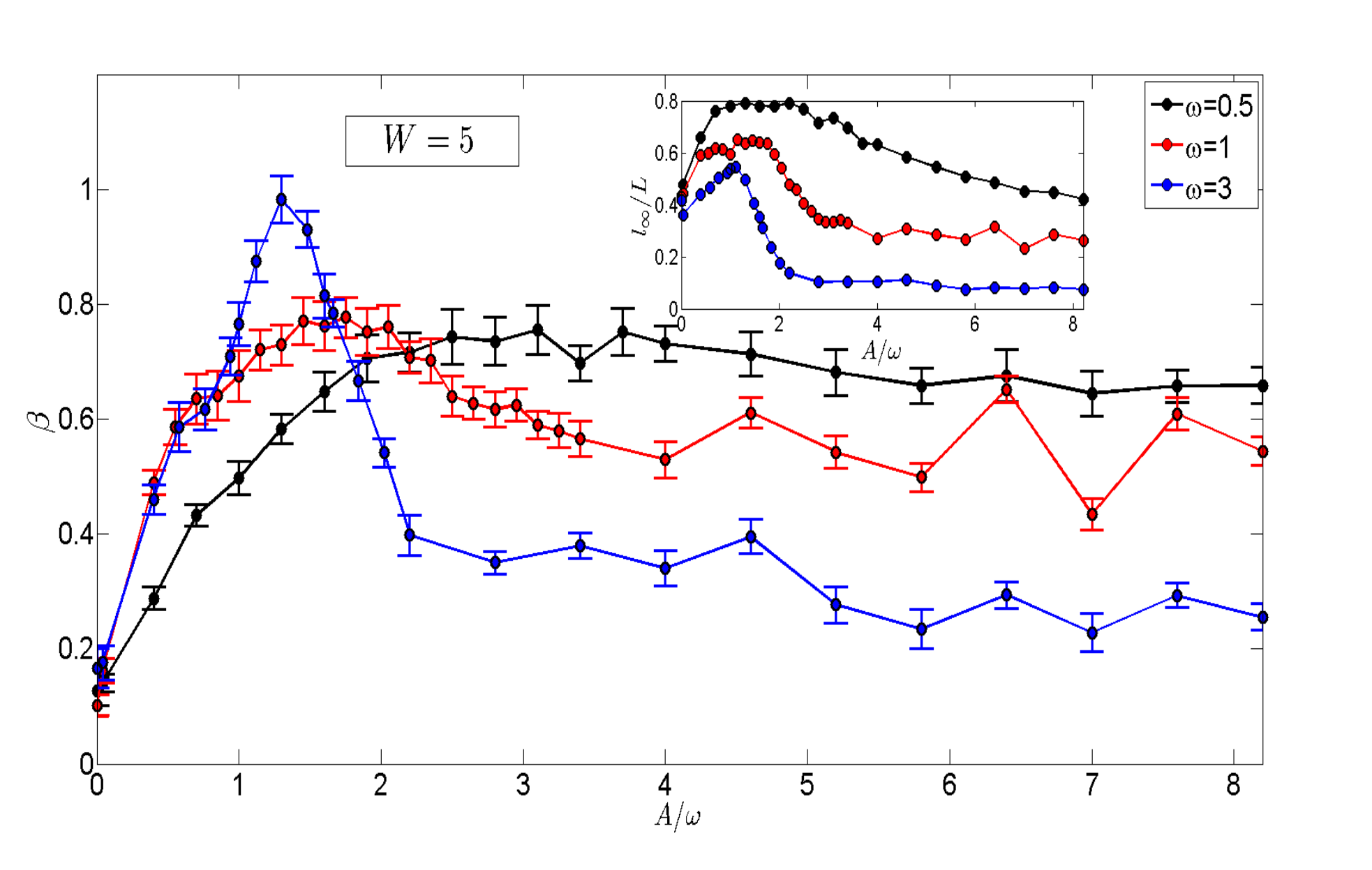}
	\caption{\label{fig:fig10} $\beta$ as a function of $A/\omega$ for different frequencies $\omega=0.5, 1, 3$. The inset shows $l_{\infty}$ as a function of $A/\omega$ for the same values of $\omega$. $E=0$ in these examples. }
\end{figure}

Looking at Eq. (\ref{eqn:floquethamiltonian2}), it is easy to realize that this system can be interpreted as a 2-D tight binding model with different longitudinal ($J$) and transverse ($A$) hopping constants, as has been discussed previously in the literature \cite{Martinez06,GomezLeon13}. Different chains correspond to different Floquet modes. The energy separation between these modes depends on the frequency of the field, so the incident electron has more effective paths to get through the system as the frequency decreases. Then, we expect the localization length to increase in the low frequency regime compared to the case without driving. However, these paths can interfere constructively or destructively, and the actual value of $l_{\infty}$ depends on the ratio $A/\omega$. In the inset of Fig.\ref{fig:fig10} we see how $l_{\infty}$ reaches a maximum at certain value of $A/\omega \sim 1$ and then starts to decrease with increasing amplitude. A more detailed discussion with a fit to a
  model with an effective number of channels has been provided by Martinez and Molina\cite{Martinez06}.

In understanding why the level repulsion parameter $\beta$ behaves in such a different way from $l_\infty$, we need to understand the structure of the quasi-energy spectrum. The quasi-energy spectrum is periodic with period $\omega$. When $\omega$ is smaller than the bandwidth of the undriven system, the spectrum has to be wrapped into itself in such a way that states coming from different parts of the spectrum can now neighbor each other. The results are then nearly Energy independent, as has been mentioned before. These different parts of the spectrum are much less correlated than neighboring parts in the original undriven systems. For $A=0$ they can be considered as having different symmetries corresponding to a different number of photons. They start mixing and repelling each other only due to the amplitude of the external field, which results in quasi-energy eigenstates with components in the different Floquet modes. The effects of this are clearly seen in the small amplitude
  results in Fig. \ref{fig:fig10}, where in all cases we start from a very small value of $\beta$ close to the Poisson limit irrespective of the original value in the undriven system. For larger amplitudes of the external field the level repulsion starts to increase, reaching a maximum at the same value of $A/\omega$ where a maximum of the localization length occurs. We see two different types of behavior. For values of the frequency $\omega$ not far from the boundary between the high and low frequency regimes ($\omega=3$ in Fig. \ref{fig:fig10}) we see a sharp peak before a slow decay to the asymptotic value. For these frequencies the localization length and the level repulsion are still correlated as seen in the curve of Fig. \ref{fig:fig9} corresponding to $\omega=3$. Deeper in the low frequency regime, we observe only a very slow decay from the maximum value of $\beta$ without a distinct peak.

\section{CONCLUDING REMARKS}
\label{sec:conclusions}

%The scaling law between $\beta$ parameter and $l_\infty$ found in the 1-D Anderson model seems not being system's specific but something related with some physical fundamental concept. 

%Numerical results suggests some relation with \textbf{dimensionality} of the system. 

A linear scaling law between the spectral repulsion parameter $\beta$ and the wave function localization length $l_{\infty}$ holds in the 1-D Anderson model.
If we include a periodic driving in the system, the same relationship holds only in the
high-frequency regime when $\omega>(W+4J)/2$. The Anderson-Floquet model in the high frequency regime behaves as the one-dimensional Anderson model without driving but with a renormalized hopping rate. This effect is caused by the coherent destruction of tunneling and has been known since the early 1990s. We have confirmed here its effects on the spectral statistics.

The low frequency regime is much richer. The different Floquet bands overlap. For small values of the field amplitude the bands almost do not hybridize, and the effect on the spectral statistics amounts to having a spectrum with mixed symmetries tending to Poisson. As the interaction between the states increases with the field strength, the spectral statistics go to the GOE limit. The localization length, on the contrary, tends to increase in the low-frequency limit, at least for not too high values of the driving field amplitude. It is then clear that the properties of the system would depend not only on the localization length of the states but also on the frequency and amplitude of the driving field. In the low-frequency regime the parameters of the external time-dependent field change the effective dimensionality of the system, as already seen for the conductance distribution \cite{Gopar10}. We lose, then, the Single Parameter Scaling fundamental to the scaling theory of Anderson localization.

\acknowledgements

We acknowledge financial support through Spanish grants MINECO No. FIS2012-34479 and MINECO/FEDER No. FIS2015-63770-P and through CAM research consortium QUITEMAD+ S2013/ICE-2801. 

\appendix

\section{$l_\infty$ vs $\beta$ in the 1-D Anderson model}
\label{sec:appendix}

\begin{figure}[!t]
	\includegraphics[width=\columnwidth, height=5cm ]{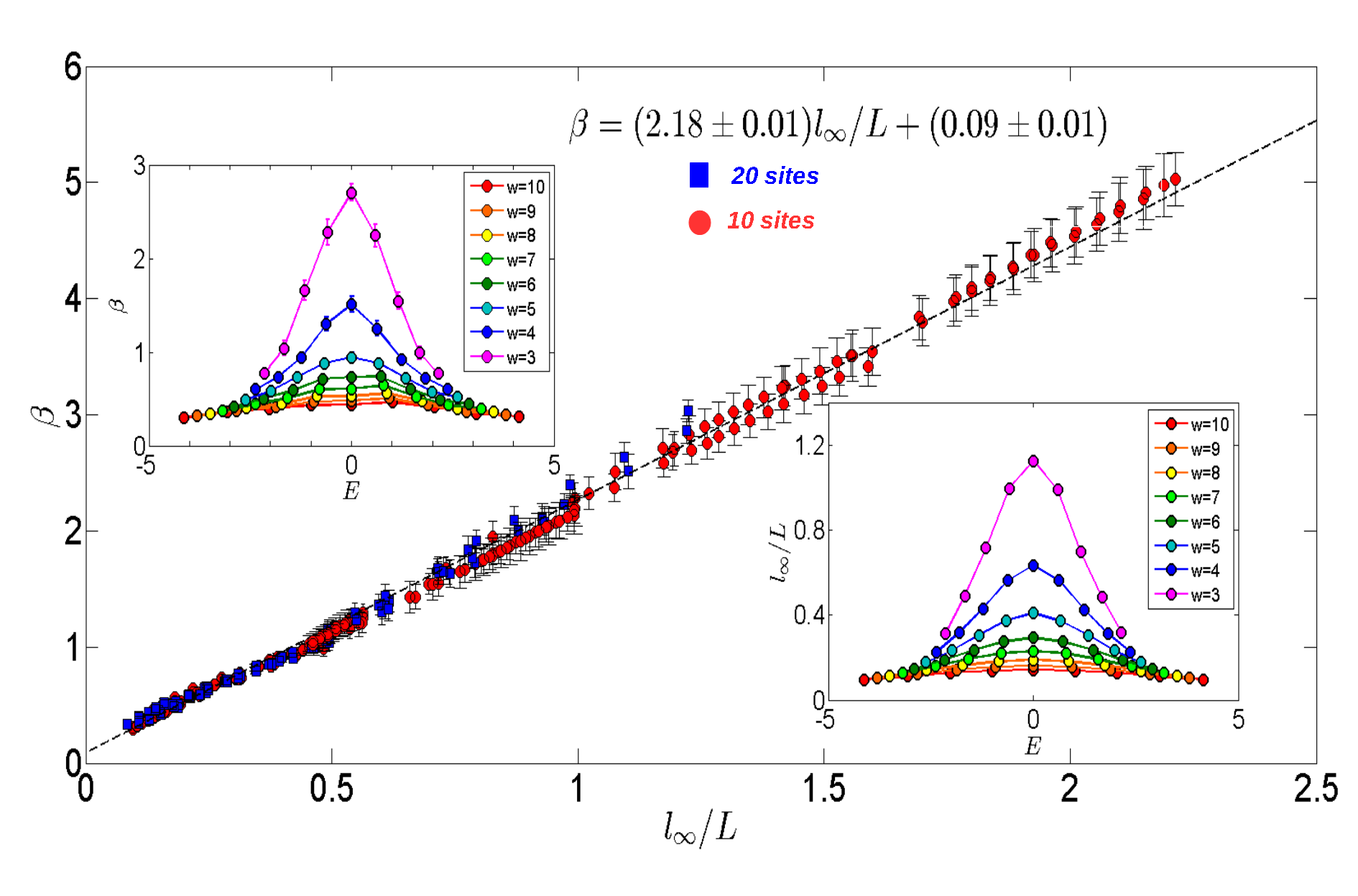}
	\caption{\label{fig:appendix}Ensemble average values of the level repulsion $\beta$ vs the ratio of the localization length and the system size $l_{\infty}/L$ for different values of disorder in the 1-D Anderson model for $10$ and $20$ sites. The insets show the energy dependence of $\beta$ and $\l_{\infty}$ for the 10 site case.}
\end{figure}	 

In this appendix we show the results obtained for the 1-D Anderson model 
%(Figs.\ref{fig:fig11}, \ref{fig:fig12}, \ref{fig:fig13}).
using the same ensemble unfolding procedure used in the main text for the Floquet-Anderson model.  
As we pointed out before, we are interested in testing the relationship between the level repulsion parameter $\beta$ and the ratio of the localization length and the length of the system $l_{\infty}/L$ in short Anderson chains. In this case, we present computations made with Anderson chains of 10 and 20 sites. 
%Usual techniques of unfolding spectra, like nearest-neighbour level spacings (see, for example \cite{haake} ) are useless with such small systems, so we are forced to use an ensemble unfolding over different realizations.
Standard techniques for unfolding the spectra are not useful for these short chains as the smooth level density is very hard to find or even define and local unfolding does not have enough levels in the window \cite{Haake_book,Gomez02}. The ratio of neighboring spacings first defined in Ref. \cite{Oganesyan07}, which is normally used without unfolding, does not work for short Anderson model chains due to the high variations of the level density for the typical energy differences of one spacing. We use instead an ensemble unfolding where we average over all the spacings $S_i=E_{i+1}-E_i$ for each value of the order index $i$ in the different realizations with the same length and disorder strength. The resulting spacing distribution is assigned to its average energy $E(S_i)=\left<E_{i+1}+E_{i}\right>/2$.

In the insets of Fig. \ref{fig:appendix} we show the results of $\beta$ vs energy and $l_{\infty}$ vs. energy for different values of the disorder strength and $L=10$. In the main panel of Fig. \ref{fig:appendix} we show the value of $\beta$ vs $l_{\infty}/L$ for $L=10$ and $L=20$; the results can be fitted with a straight line,
\begin{equation}
\beta=(2.18 \pm 0.01) \frac{l_{\infty}}{L}+(0.09 \pm 0.01)
\end{equation}
and the parameters are very close to the ones quoted by Sorathia {\em et al.} \cite{Sorathia12}, which provides justification for the use of the ensemble unfolding for Anderson models. 

%\begin{figure}[!t]
%	\includegraphics[width=\columnwidth, height=5cm ]{betavslandersonv6}
%	\caption{\label{fig:appendix}Ensemble average values of the level repulsion $\beta$ vs. the ratio of the localization length and the system size $l_{\infty}/L$ for different values of disorder in the 1-D Anderson model for $10$ and $20$ sites. The insets show figures for the energy dependence of $\beta$ and $\l_{\infty}$ in the 10 site case.}
%\end{figure}	 

%\begin{figure}[h]
%	\includegraphics[width=\columnwidth, height=5cm ]{lambdavsEanderson}
%	\caption{\label{fig:fig12}$\l_\infty$ vs Energy for different disorder realizations in the 1-D Anderson model for the same size used in the \textbf{Anderson-Floquet model}}
%\end{figure}	
%\begin{figure*}
%	\includegraphics[width=\columnwidth]{betavslambdaandersonnozero}
%	\caption{\label{fig:fig13}$\beta-l_\infty$ relation in the 1-D Anderson model for 10 and 20 sites and different values of the disorder strength.}
%\end{figure*}


\begin{thebibliography}{cc}

\bibitem{Lehmann02} J. Lehmann, S. Kohler, P. H\"anggi, A. Nitzan, Molecular wires acting as coherent quantum ratchets, Phys. Rev. Lett. \textbf{88}, 228305 (2002).

\bibitem{Platero04} G. Platero, R. Aguado, Photon-assisted transport in semiconductor nanostructures, Phys. Rep. \textbf{395}, 1 (2004).

\bibitem{Camalet04} S. Camalet, S. Kohler, P. H\"anggi, Shot noise control in ac-driven nanoscale conductors, Phys. Rev. B \textbf{70}, 155326 (2004).
  
\bibitem{Kohler05} S. Kohler, J. Lehmann, P. H\"anggi, Driven quantum transport on the nanoscale, Phys. Rep. \textbf{406}, 379 (2005).

\bibitem{Martinez08} D.F. Martinez, R.A. Molina, B. Hu, Length-dependent oscillations in the dc conductance of laser-driven quantum wires, Phys. Rev. B \textbf{78}, 045428 (2008).

\bibitem{Gu11} Z. Gu, H. A. Fertig, D. P. Arovas, and A. Auerbach,
Floquet spectrum and transport through an irradiated
graphene ribbon, Phys. Rev. Lett. \textbf{107}, 216601 (2011).

\bibitem{Lopez15} 
A. L\'opez, A. Scholz, B. Santos, and J. Schliemann, Photoinduced pseudospin effects in silicene beyond the off-resonant condition, Phys. Rev. B {\bf 91}, 125105 (2015).

\bibitem{Kirilyuk10} A. Kirilyuk, A. V. Kimel, T. Raising, Ultrafast optical manipulation of magnetic order, Rev. Mod. Phys. \textbf{82}, 2731 (2010).

\bibitem{Takayoshi14} S. Takayoshi, M. Sato, T. Oka, Laser-induced magnetization curve, Phys. Rev. B \textbf{90}, 214413 (2014).

\bibitem{Eckardt05} A. Eckardt, C. Weiss, M. Holthaus, Superfluid-insulator transition in a periodically driven optical lattice, Phys. Rev. Lett. \textbf{95}, 260404 (2005).

\bibitem{Santos09} J. Santos, R.A. Molina, J. Ortigoso, M. Rodr\'{\i}guez, Controlled localization of interacting bosons in a disordered optical lattice, Phsy. Rev. A \textbf{80}, 063602 (2009).

\bibitem{Ponte15} P. Ponte, Z. Papic, F. Huveneers, D.A. Abanin, Many-body localization in periodically driven systems, Phys. Rev. Lett. \textbf{114}, 140401 (2015).

\bibitem{Oka09} T. Oka, H. Aoki, Photovoltaic Hall effect in graphene, Phys. Rev. B \textbf{79}, 081406 (2009).

\bibitem{Kitagawa10} T. Kitagawa, E. Berg, M. Rudner, E. Demler, Topological charactarization of periodically driven quantum systems, Phys. Rev. B \textbf{82}, 235114 (2010).

\bibitem{Lindner11} N.H. Lindner, G. Refael, V. Galitski, Floquet topological insulator in semiconductor quantum wells, Nat. Phys. \textbf{7}, 490 (2011).

\bibitem{GomezLeon13} A. G\'omez-Le\'on, G. Platero, Floquet-Bloch theory and topology in periodically-driven lattices, Phys. Rev. Lett. \textbf{110}, 200403 (2013).

\bibitem{Narayan15}
A. Narayan, Floquet dynamics in two-dimensional semi-Dirac semimetals and three-dimensional Dirac semimetals, Phys. Rev. B {\bf 91}, 205445 (2015).

\bibitem{Gonzalez16} J. Gonz\'alez, R.A. Molina, Macroscopic degeneracy of zero-mode rotating surface states in 3D Dirac and Weyl semimetals, Phys. Rev. Lett. \textbf{116}, 156803 (2016).

\bibitem{Keay95} B.J. Keay, S. Zeuner, S.J. Allen Jr., K.D. Maranowski, A.C. Gossard, U. Bhattacharya, M.J.W. Rodwell, Phys. Rev. Lett. \textbf{75}, 4102 (1995).

\bibitem{Madison95} K.W. Madison, M.C. Fischer, R.B. Diener, Q. Niu, M.G. Raizen, Dynamical Bloch band suppresion in an optical lattice, Phys. Rev. Lett. \textbf{81}, 5093 (1998).

\bibitem{Guhr07} D. C. Guhr, D. Rettinger, J. Boneberg, A. Erbe, P. Leiderer, E. Scheer, Influence of laser light on electronic transport through atomic-size contacts, Phys. Rev. Lett. \textbf{76}, 086801 (2007).

\bibitem{Lignier07} H. Lignier, C. Sias, D. Ciampini, Y. Singh, A. Zenesini, O. Morsch, E. Arimondo, Phys. Rev. Lett. \textbf{99}, 220403 (2007).

\bibitem{Wang13}  Y. H. Wang, H. Steinberg, P. Jarillo-Herrero, and N.
Gedik, Observation of Floquet-Bloch states on the surface of a topological insulator, Science \textbf{342}, 453 (2013).

\bibitem{Else16} D.V. Else, B. Bauer, C. Nayak, Phys. Rev. Lett. \textbf{117}, 090402 (2016).

\bibitem{Zhang17} Z. Zhang et al., Nature \textbf{543}, 217 (2017).

\bibitem{Choi17} S. Choi et al., Nature \textbf{543}, 221(2017).

\bibitem{Dunlap86} D.H. Dunlap and V.M. Kenkre, Dynamic localization of a charged particle moving under the influence of an electric field, Phys. Rev. B \textbf{34}, 3625 (1986).


\bibitem{Grossmann91} F. Grossmann, T. Dittrich, P. Jung, P. H\"anggi, Coherent destruction of tunneling, Phys. Rev. Lett. \textbf{67}, 516 (1991).

\bibitem{Creffield07} C.E. Creffield, Quantum control and entanglement using periodic driving fields, Phys. Rev. Lett. \textbf{99}, 110501 (2007).

\bibitem{Holthaus95} M. Holthaus, G.H. Ristow, D.W. Hone, ac-Field-Controlled Anderson Localization in Disordered Semiconductor Superlattices, Phys. Rev. Lett. \textbf{75}, 3914 (1995).

\bibitem{Holthaus96} M. Holthaus, D.W. Hone, Localization effects in ac-driving tight-binding lattices, Phil. Mag. B \textbf{74}, 105 (1996).

\bibitem{Abrahams79} E. Abrahams, P.W. Anderson, D.C. Licciardello, T.V. Ramakrishnan, Scaling theory of localization: absence of quantum diffusion in two dimensions, Phys. Rev. Lett. \textbf{42}, 673(1979).

\bibitem{Bohigas84} O. Bohigas, M.J. Giannoni, C. Schmit, Characterization of chaotic quantum spectra and universality of level fluctuation laws, Phys. Rev. Lett. \textbf{52}, 1 (1984).

\bibitem{Mehta_Book} M.L. Mehta {\it Random Matrices} (Academic Press, New York, 2004).

\bibitem{Efetov83} K. B. Efetov, Supersymmetry and the theory of disordered metals, Ad. Phys. \textbf{32}, 53 (1983). 

\bibitem{Evers08} F. Evers, A.D. Mirlin, Anderson transitions, Rev. Mod. Phys. \textbf{80}, 1355 (2008).

\bibitem{Sorathia12} S. Sorathia, F.M. Izrailev, V.G. Zelevinsky, G.L. Celardo, From closed to open one-dimensional Anderson model: transport versus spectral statistics, Phys. Rev. E \textbf{86}, 011142 (2012).

\bibitem{Flores13} J. Flores, L. Guti\'errez, R.A. M\'endez-S\'anchez, G. Monsivais, P. Mora, A. Morales, Anderson localization in finite disordered vibrating rods, EPL \textbf{101}, 67002 (2013).

\bibitem{Martinez06} D.F. Martinez, R.A. Molina, Delocalization induced by low-frequency driving in disordered tight-binding lattices, Phys. Rev. B \textbf{73}, 073104 (2006).

\bibitem{Martinez06b} D.F. Martinez, R.A. Molina, Localization properties of driven disordered one-dimensional systems, Eur. Phys. J B {\bf 52}, 281 (2006).

\bibitem{Gopar10} V. A. Gopar, R. A. Molina, Controlling conductance statistics of quantum wires by driving ac fields, Phys. Rev. B \textbf{81}, 195415 (2010)

\bibitem{Kitagawa12} T. Kitagawa, T. Oka, E. Demler, Photo control of transport properties in a disordered wire: Average conductance, conductance statistics, and time-reversal symmetry, Ann. Phys. {\bf 327}, 1868 (2012).

\bibitem{Haake_book} F. Haake, \emph{Quantum signatures of chaos}, 2nd ed.,(Springer, Berlin, 2010). % check

\bibitem{Dimitriu02} I. Dimitriu, A. Edelman, Matrix models for beta ensembles, J. Math. Phys. {\bf 43}, 5830 (2002).

\bibitem{Dimitriu05} I. Dimitriu, A. Edelman, Eigenvalues of Hermite and Laguerre ensembles: large beta asymptotics, Ann. I. H. Poincaré {\bf 41}, 1083 (2005).

\bibitem{Relano08} A. Rela\~no, L. Mu\~noz, J. Retamosa, E. Faleiro, R.A. Molina, Power spectrum characterization of the continuous Gaussian ensemble, Phys. Rev. E {\bf 77}, 031103 (2008).

\bibitem{Brody73} T.A. Brody, A statistical measure for the repulsion of energy levels, Lett. Nuovo Cimento \textbf{7}, 482 (1973).

\bibitem{Izrailev88} F.M. Izrailev, Quantum localization and statistics of quasienergy spectrum in a classically chaotic system, Phys. Lett. A {\bf 134}, 13 (1988).

\bibitem{Izrailev89} F.M. Izrailev, Intermediate statistics of the quasi-energy spectrum and quantum localisation of classical chaos, J. Phys. A: Math. Gen. {\bf 22}, 865 (1989).

\bibitem{Martinez03} D. F. Martinez, Floquet-Geen function formalism for harmonically driven Hamiltonians, J. Phys. A: Math. Gen. \textbf{36}, 9827 (2003).

\bibitem{Pastawski83} H.M. Pastawski, J.F. Weisz, S. Albornoz, Matrix continued-fraction calculation of localization length, Phys. Rev. B \textbf{28}, 6896 (1983).

\bibitem{Pastawski01} E. Medina, H.M. Pastawski, Tight binding methods in quantum transport through molecules and small devices: from the coherent to the decoherent description, Rev. Mex. Fis. \textbf{47} Supp. 1, 1 (2001).

\bibitem{Gomez02} J.M.G. G\'omez, R.A. Molina, A. Rela\~no, J. Retamosa, Misleading signatures of chaos, Phys. Rev. E \textbf{66}, 036209 (2002). 

\bibitem{Oganesyan07} V. Oganesyan, D.A. Huse, Localiztion of interacting fermions at high temperature, Phys. Rev. B \textbf{75}, 155111 (2007).

\bibitem{Thouless79} D.J. Thouless, in \emph{Ill-Condensed Matter}, edited by R. Balian, R. Maynard and G. Toulouse (North-Holland, Amsterdam, 1979).

\bibitem{Altshuler03} L.I. Deych, M. V. Erementchouk, A. A. Lisyansky, and B. L. Altshuler, Scaling and the Center-of-Band Anomaly in a One-Dimensional Anderson Model with Diagonal Disorder, Phys. Rev. Lett. \textbf{91} 096601 (2003).

\bibitem{Porter65} C.E. Porter, ed., \emph{Statistical Theories of Spectra: Fluctuations}, Academic Press, New York, (1965) 

\bibitem{Brody81} T.A. Brody, J. Flores, J.B. French, P.A. Mello, A. Pandey and S.S.M. Wong, Random-matrix physics: spectrum and strength fluctuations, Rev. Mod. Phys. \textbf{53}, 385 (1981) 

\bibitem{Kramer93} B. Kramer and A. MacKinnon, Localization: theory and experiment, Rep. Prog. Phys. \textbf{56}, 1469 (1993).

\bibitem{Izrailev97}F. M. Izrailev, T. Kottos,  A. Politi, and G. P. Tsironis, Evolution of wave packets in quasi-one-dimensional and one-dimensional random media:
Diffusion versus localization, Phys. Rev. E. \textbf{55}, 4951 (1997)

\end{thebibliography}
\end{document}